\documentstyle[12pt]{article}
\begin{document}
\title{Dark Energy, Gravitation and Electromagnetism}
\author{B.G. Sidharth\\
International Institute of Applicable Mathematics \& Information Sciences\\
Hyderabad (India) \& Udine (Italy)\\
B.M. Birla Science Centre, Adarshnagar, Hyderabad - 500 063,
India}
\date{}
\maketitle
\begin{abstract}
In the context of the fact that the existence of dark energy causing the accelerated expansion of the universe has been confirmed by the WMAP and the Sloan Digital Sky Survey, we re-examine gravitation itself, starting with the formulation of Sakharov and show that it is possible to obtain gravitation in terms of the electromagnetic charge of elementary particles, once the ZPF and its effects at the Compton scale are taken into account.
\end{abstract}
\section{Introduction}
Some years ago the author had put forward a model in which dark energy in the form of a background Zero Point Field would lead to an accelerating and expanding universe \cite{bgs1,bgs2,bgs3,bgs4}. Subsequently observations of distant Type 1A Supernovae by Perlmutter, Schmidt and others confirmed that the universe was indeed accelerating and expanding \cite{perl,kirsh}. However final confirmation for the dark energy contribution to this observed acceleration and expansion was awaited. This confirming evidence has now been obtained through the observations of the Wilkinson Microwave Anisotropy Probe and the Sloan Digital Sky Survey \cite{science}. In the light of the above observational evidence, we will now try to reexamine and characterize gravitation and electromagnetism.
\section{Gravitation}
Our starting point is the 1967 analysis of Sakharov \cite{sakharov,mwt}. He starts with the space time and curvature action integral of General Relativity
\begin{equation}
S (R) = - \frac{1}{16\pi G} \int (dx) \sqrt{-g}R\label{e1}
\end{equation}
He then observes that the action (\ref{e1}) implies a metrical elasticity of space, that is generalized forces opposing the curving of space. He then goes on to the suggestion of Zeldovich \cite{zeldovich}, ``... the gravitational interactions could lead to a ``small'' dusturbance of this equilibrium and thus to a finite value of Einstein's cosmological constant, in agreement with the recent interpretation of the astrophysical data.''\\
Expanding the Lagrangian density function in (\ref{e1}) in a power series in $R$, the scalar curvature, we get
\begin{equation}
L (R) = L(0) + A \int k dk \cdot R + B \int \frac{dk}{k} R^2 + \cdots\label{e2}
\end{equation}
where $A$ and $B$ are of the order of $1$.\\
Sakharov then identifies the universal constant of gravitation from the second term in (\ref{e2}) as
\begin{equation}
G = -\frac{1}{16\pi A \int k dk}, \quad A \sim 1\label{e3}
\end{equation}
There is a  divergence in the denominator of (\ref{e3}). To circumvent this Sakharov postulates a minimum length cut off at the Planck scale, that is a cut off at the Planck mass, $10^{-5}gm$.\\
If we implement this cut off in (\ref{e3}) we get
\begin{equation}
G m^2_P \sim 1\label{e4}
\end{equation}
The importance of the above considerations is that they show that, ``...the magnitude of the gravitational interaction is determined by the masses and equations of motion of free particles, and also, probably, by the momentum cut off.'' More explicitly the above analysis demotes gravitation to the analogue of elasticity in Chemical Physics, that is a mere statistical measure of residual energies (Cf.ref.\cite{mwt}). (We will further justify this in this sequel). On the other hand it must also be borne in mind that the Planck mass particle is not an elementary particle. It would  at the scale of Planck dimensions decay in $10^{-42}$seconds. Moreover Rosen has shown that it would constitute a mini universe in itself \cite{rosen}.\\
Nevertheless it is possible to extend the above arguments and get results in the observable realm of elementary particles by considering the background Zero Point Field (ZPF) or Dark Energy alluded to earlier (Cf.ref.\cite{cu,lee,miloni,iz}.\\
We note that as is well known, such a background ZPF can explain the Quantum Mechanical spin half as also the anomalous $g = 2$ factor for an otherwise purely classical electron \cite{sachi,boyer}. The key point here is (Cf.ref.\cite{sachi}) that the classical angular momentum $\vec r \times m \vec v$ does not satisfy the Quantum Mechanical  commutation rule for the angular momentum $\vec J$. However when we introduce the background Zero Point Field, the momentum now becomes
\begin{equation}
\vec J = \vec r \times m=\vec v + (e/2c) \vec r \times (\vec B \times \vec r) + (e/c) \vec r \times \vec A^0 ,\label{e5}
\end{equation}
where $\vec A^0$ is the vector potential associated with the ZPF and $\vec B$ is an external magnetic field introduced merely for convenience, and which can be made vanishingly small.\\
It can be shown that $\vec J$ in (\ref{e5}) satisfies the Quantum Mechanical commutation relation for $\vec J \times \vec J$. At the same time we can deduce from (\ref{e5})
\begin{equation}
\langle J_z \rangle = - \frac{1}{2} \hbar \omega_0/|\omega_0|\label{e6}
\end{equation}
Relation (\ref{e6}) gives the correct Quantum Mechanical results referred to above.\\
From (\ref{e5}) we can also deduce that
\begin{equation}
l = \langle r^2 \rangle^{\frac{1}{2}} = \left(\frac{\hbar}{mc}\right)\label{e7}
\end{equation}
(\ref{e7}) shows that the mean dimension of the region in which the fluctuation contributes is of the order of the Compton wavelength of the electron.\\
As a further confirmation, we can similarly deduce that there is a mean time interval of the order of the electrons' Compton time. For this we note that the energy of the ZPF is given by \cite{boyer}
$$\frac{1}{8\pi}\langle E^2 + B^2 \rangle = \frac{1}{8\pi} \sum^{2}_{\lambda = 1} \int d^2 kh^2 (\omega_\hbar)$$
\begin{equation}
= \int^\infty_{k=0} dk k^2h^2(\omega_\hbar) = \int^\infty_{\omega = 0} d\omega^{\omega^2}_{c^3}-h^2(\omega),\label{e8}
\end{equation}
where the spectral energy density is given by
\begin{equation}
\rho (\omega) =  \frac{\hbar}{2\pi^2} \frac{\omega^3}{c^3}\label{e9}
\end{equation}
If now we denote the impulsive change of momentum in the interval $\tau$ due to the buffetting of the particle by the ZPF by 
$$\langle \Delta^2 \rangle^{\frac{1}{2}}$$
then we can easily deduce that by (\ref{e8}) and (\ref{e9}), 
\begin{equation}
\langle \Delta^2 \rangle = \left(r \Gamma \pi^4 c^4 \tau/5\omega^2 \right) \cdot \rho^2 (\omega ,T)\label{e10}
\end{equation}
where
\begin{equation}
\rho (\omega , T) = \left(\frac{\omega^2}{2\pi^2c^3}\right) \langle m v^2 \rangle\label{e11}
\end{equation}
In (\ref{e11}) $v$ is the root mean square velocity of the rapidly vibrating particle, which we take to be $c$, the velocity of light. Then we can deduce from (\ref{e10}), on using the fact that the magnitude of the impulsive change of the momentum would be $mc$, that
\begin{equation}
\tau \sim \frac{\hbar}{mc^2}\label{e12}
\end{equation}
(\ref{e12}) shows that $\tau$ of the order of the Compton time.\\
If we use the fact that there is the Compton time uncertainty $\sim \tau$ in (\ref{e3}) and replace 
$$G = G \left(t\right)$$
by
$$G\left(t - \tau\right)$$
and further use the fact to be elaborated that
\begin{equation}
\dot {G} = - G/t\label{e13}
\end{equation}
We can easily see that the factor $A$ acquires an extra term  $\frac{A t}{\tau}$. This immediately leads to the length cut off being
\begin{equation}
l = l_P \sqrt{\frac{t}{\tau}}\label{e14}
\end{equation}
(\ref{e14}) shows that over and above the cut off in (\ref{e3}) and the Planck mass in (\ref{e4}), there is due to the Compton scale average a further contribution at the cut off length $l$ which is the Compton wavelength of an elementary particle, the corresponding mass being the mass of an elementary particle.\\
(\ref{e13}) is well known \cite{nar,cu,fpl}. Its significance in the above context will be seen below. If we now use the extra contribution to (\ref{e4}) with (\ref{e14}) we deduce the well known otherwise, empirical electromagnetism - gravitation coupling strengths ratio
\begin{equation}
\frac{Gm^2}{e^2} \sim 10^{-40}\label{e15}
\end{equation}
In other words the extra contribution to the action integrals in (\ref{e1}) or (\ref{e2}) from the Compton scale ZPF effects leads to a characterization of a gravitational constant in terms of the electric charge. Looking at it another way, this points to a unified description of gravitation and electromagnetism, because (\ref{e15}) is no longer ad hoc or empirical, but rather is a consequence of the theory.\\
Let us now look at how relation (\ref{e13}) can be arrived at from the point of view of fluctuations in the number of particles $N \sim 10^{80}$ in the large scale universe. Due to this fluctuation which is of the order $\sqrt{N}$, the excess electromagnetic energy is given by the expression
$$\frac{e^2\sqrt{N}}{R}$$
If we now use the well known relation (Cf.ref.\cite{cu,hayakawa,nottale})
$$R = GM/c^2 = \frac{GmN}{c^2}$$
in the above, the excess energy comes out as
\begin{equation}
\mbox{Excess} = \frac{e^2c^2}{Gm\sqrt{N}}\label{e16}
\end{equation}
$M$ is the mass of the universe which equals $Nm, m$ being the mass of a typical elementary particle.
(\ref{e16}) is the contribution due to the fluctuation in the number of particles. The question is can this energy due to the particle number fluctuation be identified with the ZPF energy within the Compton scale? Infact this latter energy is given by (Cf.ref.\cite{boyer}), independently, 
\begin{equation}
\frac{1}{8\pi} \langle E^2 + B^2 \rangle = \int_\omega \frac{\hbar \omega^3}{c^3} d\omega \sim \frac{\hbar \omega^4}{c^3}\label{e17}
\end{equation}
(Cf.ref.\cite{mwt} for details). The ZPF energy (of oscillators) over an extension $l$ is given by the expression
$$\frac{\hbar c}{l^4}$$
This is seen to be the same as (\ref{e17}) because,
$$\omega \sim \frac{1}{\tau} \mbox{and} c \tau \sim l$$
So the energy in a volume is $\sim l^3$ is from the above
\begin{equation}
\mbox{Excess} = \frac{\hbar c}{l}\label{e18}
\end{equation}
which can be seen to be of the same order as (\ref{e16}) or its predecessor.\\
In other words the fluctuating energy of the ZPF at the Compton scale gives the entire energy of the elementary particle, on the one hand. On the other hand this equality of (\ref{e16}) and (\ref{e18}) shows that
\begin{equation}
G = \frac{lc^2}{m\sqrt{N}} = \frac{l^2c\tau}{mt}\label{e19}
\end{equation}
where $t$ is the age of the universe. (\ref{e19}) immediately leads to (\ref{e13}). 
\section{Discussion}
1. We have seen that the ZPF can be associated with the fluctuation in the number of particles in the considerations leading to (\ref{e16}). Another way of looking at Sakharov's formulation in (\ref{e2}) and the following considerations is that if we write the Lagrangian in (\ref{e2}) for the $\sqrt{N}$ particles, then $A$ gets multiplied by this factor and instead of (\ref{e4}), we will get, as before, (\ref{e15}). Indeed the fact that the two approaches lead to identical conclusions is due to the fact that $t = \sqrt{N} \tau$, a relation which infact can be deduced from the theory (Cf.ref.\cite{bgs1}), and the relation which is equivalent to the well known Eddington formula $r = \sqrt{N} l$.\\
2. It is interesting that the ZPF effects as discussed in equations (\ref{e5}), \ref{e7}), (\ref{e8}), (\ref{e12}) and (\ref{e17}) bear a strong resemblance to similar zitterbewegung effects at the same scale in the Quantum Mechanical treatment of an electron, and can be identified with the latter. This is similar to the case of the ZPF at the scale of the atomic radius giving rise to the Lamb Shift.\\
3. The considerations leading to (\ref{e7}) or (\ref{e12}) via (\ref{e8}) and (\ref{e10}), as also the equation (\ref{e17}) point to the fact that it is possible to consider the mass of an elementary particle being generated by the ZPF at the Compton scale. This has been discussed in detail in (Cf.ref.\cite{ijpap}).\\
4. The significance of the ZPF considerations vis-a-vis the curvature of space is the followintg: It is well known that the fluctuation in the scalar curvature in a region of extent $\lambda$ is given by
\begin{equation}
\Delta R \sim \frac{l_P}{\lambda^3}\label{e20}
\end{equation}
where $l_P$ is the Planck length $\sim 10^{-33}cms$ (Cf.ref.\cite{mwt}). Usually this quantity is miniscule. When $\lambda = l$ the Compton wavelength of for example is  electron, then it can be seen from (\ref{e20}) that $\Delta R$ is not only not small but infact is of the order $1$. Thus we have a huge fluctuation at the Compton scale.\\
5. It may be mentioned that the fluctuation $\sim \sqrt{N}$ referred to above, has been shown to be related via the order parameter to a Landau-Ginsburg type Phase transition (from the Quantum vacuum) \cite{newcosmos,sf}.\\
6. In the context of the minimum Compton scale it would be interesting to consider a modified Coulomb potential proportional to $\frac{1}{x-l}$, where $x$ is the radial distance from the origin. In this case we consider as in the usual theory, the Schrodinger equation which now becomes
\begin{equation}
u'' + \left[ k^2 - U(x)\right] u = 0,\label{e21}
\end{equation}
where $U(r) = \frac{\beta}{x-l}$\\
For $x < < l, U$ in (\ref{e21}) has the expansion
$$-\frac{\beta}{l} \left(1 + \frac{x}{l}\right) = const - \frac{\beta x}{l^2}$$
This potential agrees with the confining quark potential \cite{lee}.\\
Furthermore the wave function solution of (\ref{e21}) is now expressible in terms of the well known Airy function, and infact in this region can be approximated by
$$u \propto \exp \left(-\gamma x^{3/2}\right),$$
$$\gamma \, \mbox{a \, constant}$$
This shows that the wave function and therefore the particle is confined to a region roughly of the order of the Compton wavelength $l$ itself. We thus get a rationale for quark confinement. For the region $x$ much greater than $l$, the potential in (\ref{e21}) behaves like the normal Coulumb potential.\\
7. Following the above remarks, it is interesting to note the following. Very much in this spirit, it was argued that \cite{ffp1} a neutrino has a small electrical charge
$$e_\nu \sim 10^{-6} e,$$
$e$ is the electron charge.\\
This work infact predicted that the neutrino would have a mass, a hundred millionth that of the electron, as was subsequently confirmed by the superkamiokande experiment (Cf. also. \cite{cu}). If we now consider a tight neutrino anti-neutrino bound state, as was also argued for the pion, we get
$$e^2_\nu / r^2 = m_\nu c^2/r$$
which leads to the correct quark Compton wavelength and mass which $\sim 10^4$ times the electron mass.\\
8. In the context of the minimum Compton length considerations, it was argued that this would lead to a non-integrable derivative
$$
\frac{\partial a^\mu}{\partial x^\nu} \to \frac{\partial a^\mu}{\partial x^\nu} - \Gamma^\mu_{\lambda \nu} a^\lambda$$ 
\begin{equation}
= \frac{\partial a^\mu}{\partial x^\nu} - \Gamma^\mu_{\lambda \nu} \delta^\lambda_\rho a^\rho\label{e22}
\end{equation}
If further we have
$$a^\mu = \partial^\mu Q$$
from here we can show that the net result is that there is a covariant derivative given by
$$\partial_\nu \partial^\mu \to \partial_\nu\partial^\mu - \Gamma^\mu_{\lambda \nu} \partial^\lambda$$
whence,
\begin{equation}
\partial^\mu \to \partial^\mu - \Gamma^{\mu \nu}_\nu\label{e23}
\end{equation}
The equation (\ref{e23}) shows the emergence of the electromagnetic potential, as has been discussed in detail (Cf.ref.\cite{annales,nc117}).\\
9. A problem which has vexed Physicists is the matter anti-matter asymmetry. It may be pointed out that in the light of the above considerations, it is possible to give an argument from probability theory itself. Let us suppose that the probability for the creation of particles and anti-particles is the same, as for example the probability for giving birth to a boy or a girl. However in this latter case, the probability that any family with several children has exactly half the number of boys is infact less than half. A similar argument would apply to matter and anti-matter.

\end{document}